\documentclass[pra, aps, twocolumn, groupedaddress, showpacs, superscriptaddress]{revtex4-2}

\usepackage{amssymb,amsmath,amsthm, color,graphicx, times,graphicx}
\usepackage{hyperref}
\usepackage{enumerate}
\usepackage{subfigure}
\usepackage{times}
\usepackage{txfonts}
\usepackage{bbold}
\usepackage{color}
\usepackage{amsmath}
\usepackage{booktabs}
\usepackage{hyperref}
\usepackage{overpic}
\usepackage{algorithm}
\usepackage{algpseudocode}
\hypersetup{
    colorlinks=true,
    linkcolor=black,    
    citecolor=black,    
    urlcolor=blue       
}

\providecommand{\openone}{\leavevmode\hbox{\small1\kern-4.3pt\normalsize1}}

\hypersetup{
    colorlinks=true,
    linkcolor=green,    
    citecolor=red,    
    urlcolor=blue       
}

\theoremstyle{plain}

\usepackage{orcidlink}

\theoremstyle{definition}

\begin{document}
\title{Distributed quantum phase sensing using an SU(1,1) Perelomov coherent state in a Mach--Zehnder interferometer network}
\author{Seloua Boulifa \orcidlink{0009-0005-1979-6680}}\affiliation{LPHE-Modeling and Simulation, Faculty of Sciences, Mohammed V University in Rabat, Rabat, Morocco}
\author{Abdallah Slaoui \orcidlink{0000-0002-5284-3240}}
\email{abdallah.slaoui@um5s.net.ma}
\affiliation{LPHE-Modeling and Simulation, Faculty of Sciences, Mohammed V University in Rabat, Rabat, Morocco} \affiliation{Centre of Physics and Mathematics, CPM, Faculty of Sciences, Mohammed V University in Rabat, Rabat, Morocco}
\author{Hanane EL Hadfi}\affiliation{LPHE-Modeling and Simulation, Faculty of Sciences, Mohammed V University in Rabat, Rabat, Morocco}
\author{Rachid Ahl Laamara \orcidlink{0000-0001-8410-9983}}\affiliation{LPHE-Modeling and Simulation, Faculty of Sciences, Mohammed V University in Rabat, Rabat, Morocco} \affiliation{Centre of Physics and Mathematics, CPM, Faculty of Sciences, Mohammed V University in Rabat, Rabat, Morocco}

\begin{abstract}  
We present a theoretical study of distributed quantum sensing using a reconfigurable network of $N$ Mach--Zehnder interferometers to simultaneously estimate weighted linear combinations of relative phase shifts. The proposed scheme relies on injecting a Perelomov coherent state, generated from the SU(1,1) group representation, into one port of a passive quantum circuit while the remaining modes are initialized in the vacuum state, thereby establishing non-trivial quantum correlations among the sensors. Based on multi-parameter quantum estimation theory, our analysis shows that this entangled strategy outperforms the standard shot-noise limit and asymptotically achieves the Heisenberg limit. Furthermore, we highlight the crucial role of the first beam splitter's asymmetry; specifically, the optimal transmission coefficient that maximizes sensitivity corresponds to its maximum imbalance. Compared to a separable approach utilizing the same total average photon number, the entangled strategy yields a more efficient use of quantum resources, guaranteeing an accuracy gain between $1$ and $N$. This demonstrates that the entangled strategy is systematically superior to its separable counterpart. Ultimately, these results illustrate the potential of entangled distributed quantum sensor networks for high-precision measurements and provide clear guidelines for optimizing interferometric architectures in quantum metrology.
 
\par
\vspace{0.25cm}


\end{abstract}
\date{\today}

\maketitle

\section{Introduction}
Quantum sensing is now among the cornerstones of next-generation quantum technologies~\cite{Giovannetti2004, Giovannetti2011, Degen2017}. It aims to exploit the non-classical properties of quantum systems to improve measurement accuracy beyond the limits imposed by classical systems. In particular, the use of phenomena such as quantum superposition ~\cite{Agarwal2012}, entanglement~\cite{Horodecki2009, Pezze2009}, and quantum squeezing~\cite{Caves1981, Schnabel2017} makes it possible to
increase the sensitivity of measurement devices and approach fundamental limits, such as the Heisenberg limit, thereby exceeding the shot-noise limit~\cite{Holland1993, Giovannetti2006, Lee2002, Aasi2013}. These advances have led to the development of several technological applications, especially high-precision atomic clocks~\cite{Ludlow2015, Chou2010}, quantum magnetic-field sensors~\cite{Budker2007, Kominis2003}, inertial sensors based on atomic interferometry~\cite{Kasevich1991, Fixler2007}, as well as quantum imaging techniques~\cite{Genovese2016, Kolobov2007}. Among the most widely used experimental setups in quantum metrology, the Mach--Zehnder interferometer (MZI) plays a key role due to its ability to perform highly sensitive phase-shift measurements ~\cite{Campos1989, Abdellaoui2024, Abouelkhir2025, Yurke1986, Gerry2005}. Numerous studies have shown that the phase-estimation performance of the MZI can be enhanced by employing nonclassical input states, including Fock states, squeezed states, entangled states, and NOON states ~\cite{Pezze2013, Pezze2018, Dowling2008, Afek2010, Pezze2008, Lang2013}. Nearly optimal measurements using coherent and squeezed-vacuum inputs have also been investigated in noisy MZIs~\cite{Gard2017}, while enhanced phase sensitivity has been demonstrated using squeezed Kerr states~\cite{Yadav2024}.
Beyond single-mode nonclassical resources, two-mode squeezed states have also been extensively investigated for quantum phase estimation. In particular, the two-mode squeezed vacuum (TMSV) state has been employed in Mach--Zehnder interferometry, where its two-mode photon correlations, combined with parity detection, provide enhanced phase sensitivity and allow the corresponding quantum Cramér-Rao bound to be attained under the considered conditions~\cite{Anisimov2010}.
More general two-mode squeezed coherent (TMSC) states have also been considered for MZI-based phase estimation, where the combination of coherent displacement and two-mode squeezing enhances the phase-estimation performance, with the achievable precision analyzed through the quantum Fisher information and specific detection
schemes~\cite{Liu2023}. In parallel, SU(1,1) interferometers, in which nonlinear parametric processes replace the conventional beam splitters, have been investigated with different nonclassical input configurations, including coherent and squeezed-vacuum states~\cite{Gong2017}. More recently, TMSC states have been investigated in SU(1,1) interferometers, where the corresponding QFI and QCRB have been analyzed together with balanced-homodyne and intensity-detection schemes, including the effects of photon losses~\cite{Xu2026}. These developments highlight the metrological relevance of squeezed quantum resources and show that the achievable phase sensitivity depends on the input quantum state, the interferometric architecture, and the measurement scheme.
 
In this context, the development of networks of interconnected interferometers has recently attracted growing interest, particularly in the context of distributed quantum sensing, which involves using multiple spatially separated quantum sensors to simultaneously estimate several physical parameters or a linear combination of these parameters~\cite {Eldredge2018, PezzeSmerzi2025, Proctor2018, Ge2018, Zhuang2019, Humphreys2013}.

Distributed quantum sensor networks offer several potential advantages over individual sensors. In particular, the use of entangled states distributed across different sensors can improve the overall accuracy of the estimate and reduce the resources required to achieve a given level of sensitivity~\cite{Gessner2018, Xia2020}. These systems are particularly promising for applications involving large-scale collective measurements, such as quantum clock networks~\cite{Komar2014} and quantum navigation~\cite{Li2026}. However, the optimization of these networks and the comparison between entangled and separable sensors remain open questions and are the subject of extensive theoretical research~\cite{Pirandola2018,  Malitesta2023}. The fundamental analysis of the performance of these measurement protocols relies on tools from quantum estimation theory~\cite{Helstrom1969, Holevo2011}. In this context, the maximum achievable precision in estimating a physical parameter is limited by the quantum Cramér-Rao bound (QCRB), thereby setting a lower bound on the variance of any unbiased estimator~\cite{Braunstein1994}. This bound is directly related to the quantum Fisher Information (QFI), a quantity that measures how sensitively a quantum state reacts to infinitesimal changes in the parameter to be estimated~\cite{Toth2014, Teklu2010, Cavazzoni2025,Asjad2023}. In the context of estimating several parameters simultaneously, the limit of achievable precision is characterized by the quantum Fisher information matrix (QFIM), which generalizes QFI to the multiparameter setting and serves as the primary theoretical tool for analyzing the performance of quantum sensor networks.

Recent studies have explored the joint estimation of successive squeezing parameters, the saturation of multiparameter quantum limits in distributed quantum sensing, and multiple-phase estimation using nonclassical multimode quantum states, further highlighting the relevance of nonclassical resources in high-precision multiparameter quantum sensing ~\cite{Sharma2025, Liu2020, Safranek2018, Teklu2025, Saidi2026}. In this study, we investigate a distributed quantum sensing approach based on a configurable network of $N$ Mach--Zehnder interferometers for estimating a linear combination of relative phase shifts. A central ingredient of the proposed protocol is the choice of the input quantum resource, namely a Perelomov coherent state constructed from the SU(1,1) group~\cite{Perelomov1977,wang2000}. In the entangled strategy, a single SU(1,1) Perelomov coherent state is injected into a passive quantum circuit and coherently distributed across the $N$ sensing modes, while the remaining input modes are initialized in the vacuum state. For comparison, we consider a separable strategy in which independent SU(1,1) Perelomov coherent states are employed at the individual sensors. The two strategies are analyzed and optimized within the QFIM-QCRB framework. Their performance is compared under the same total average photon-number constraint for different linear combinations of the relative phase shifts, with the relative advantage quantified through the sensitivity gain. The main objective of this work is to investigate the metrological performance of the SU(1,1) Perelomov coherent state as a nonclassical resource for distributed phase sensing and to determine the conditions under which the entangled strategy provides an advantage over its separable counterpart.

\section{DISTRIBUTED QUANTUM SENSING}

 In distributed quantum sensing, a set of $N$ unknown parameters $ \boldsymbol{\phi}=\{\phi_1 \ldots \phi_N \}$ is encoded in distinct modes or interferometers, which may be spatially separated. The encoding of the parameters is represented by commutative unitary transformations. In the noiseless scenario, the evolution is described by: 
 \begin{equation}
e^{-i \boldsymbol{\phi \cdot  \hat{\mathcal{H}}}} = \bigotimes_{k=1}^N e^{-i  \hat{\mathcal{H}}_k \phi_k},
\end{equation}
where $\boldsymbol{\hat{\mathcal{H}}} = \{\hat{\mathcal{H}}_1, \ldots, \hat{\mathcal{H}}_N\}$ is a set of commutative Hermitian operators such that $ [\hat{\mathcal{H}}_k, \hat{\mathcal{H}}_l] = 0$ for all $k,l = 1,\ldots,N$, and $\boldsymbol{\phi \cdot \hat{\mathcal{H}}}= \sum_{k=1}^N \phi_k \hat{\mathcal{H}}_k$. The sensing protocol aims to estimate a linear combination \cite{liu2021, pezze2015, gross2021} \begin{equation} 
    \boldsymbol{\mu} \cdot \boldsymbol{\phi} = \sum_{k=1}^{N} \mu_{k} \phi_{k},
\end{equation}

with the normalization condition $\sum_{k=1}^{N} \mu_{k}^{2} = 1/N$.  For this, two strategies can be distinguished. The first corresponds to an \textit{entangled strategy} (also referred to as \textit{global} or \textit{parallel}), in which the overall state $\hat{\rho}$ of the network is prepared in an entangled state of modes. The second corresponds to a \textit{separable strategy} (or \textit{local} / \textit{sequential}), where the probe state takes the factorized form
\begin{equation}
\hat{\rho} = \bigotimes_{k=1}^N \hat{\rho}_k,
\end{equation}
with $\hat{\rho}_k$ the state associated with the $k$-th sensor. In this case, the sensors operate without quantum correlation, and the parameters $\phi_1, \ldots, \phi_N$ are estimated independently. The performance of the estimation protocol is quantified by a fundamental
quantity:
\begin{equation}
    \Delta^{2}(\boldsymbol{\mu}\cdot\boldsymbol{\phi})_{\mathrm{QCRB}}
    =
    \boldsymbol{\mu}^{T}\boldsymbol{F}_{Q}^{-1}\boldsymbol{\mu},
    \label{QCRB}
\end{equation}
where $\boldsymbol{F}_{Q}$ denotes the QFIM, of dimension $N\times N$, which characterizes the quantum sensitivity of the estimation protocol. For a pure probe state $\hat{\rho}=|\psi\rangle\langle\psi|$, the elements of the QFIM take the form
\begin{equation}
(\boldsymbol{F}_Q)_{kl}
=
4\,\mathfrak{R}\Big\{
\langle\partial_k\psi|\partial_l\psi\rangle
-
\langle\partial_k\psi|\psi\rangle
\langle\psi|\partial_l\psi\rangle
\Big\},
\label{FQ}
\end{equation}
where $\partial_k\equiv\partial/\partial\phi_k$.

The mode-entangled strategy considered below is characterized by multimode correlations induced by the quantum circuit, which generally give rise to nonzero off-diagonal elements of the QFIM. In this case, we denote the corresponding QCRB by $\Delta^{2}(\boldsymbol{\mu}\cdot\boldsymbol{\phi})_{\mathrm{QCRB}}^{\,\mathrm{me}}$, where the superscript 'me' refers to the mode-entangled strategy. In contrast, the mode-separable strategy is described by a tensor-product
probe state.
\begin{equation}
|\psi\rangle
=
\bigotimes_{k=1}^{N}|\psi_k\rangle.
\end{equation}
In this case, the QFIM is diagonal, and we denote the corresponding QCRB by $\Delta^{2}(\boldsymbol{\mu}\cdot\boldsymbol{\phi})_{\mathrm{QCRB}}^{\,\mathrm{ms}}$,
where the superscript 'ms' refers to the mode-separable strategy. Consequently, Eq.~\eqref{QCRB} reduces to
\begin{equation}
\Delta^{2}(\boldsymbol{\mu}\cdot\boldsymbol{\phi})_{\mathrm{QCRB}}^{\,\mathrm{ms}}
=
\sum_{k=1}^{N}\frac{\mu_k^{2}}{F_{Q_k}},
\label{QCRBs}
\end{equation}
where $F_{Q_k}=(\boldsymbol{F}_{Q})_{kk}$. Its diagonal elements are given
by
\begin{equation}
F_{Q_k}
=
4\left(
\langle\psi_k|\hat{\mathcal{H}}_k^2|\psi_k\rangle
-
\langle\psi_k|\hat{\mathcal{H}}_k|\psi_k\rangle^2
\right),
\label{FQs}
\end{equation}
where $\hat{\mathcal{H}}_k$ is the generator associated with the phase
parameter $\phi_k$.
In non-Bayesian estimation, $F_{Q}$, as a measure of information, is additive~\cite{Paris2009}, indicating that repeating the same experiment $M$ times corresponds to multiplying the QFI by $M$, i.e., $F_{Q} \rightarrow M F_{Q}$; as a result, the limits indicated in equations \eqref{QCRB} and \eqref{QCRBs} naturally take into account the number $M$ of independent repeated measurements.
In this work, this factor is neglected by considering $M=1$, which simplifies the expressions while preserving the fundamental analysis. For a detailed discussion of multiparametric Bayesian estimation explicitly including $M$ as a resource, see reference~\cite{Gebhart2021, Teklu2009}.
\section{Distributed Mach–Zehnder Sensing}
\begin{figure}[htbp]
     \begin{minipage}[b]{0.6\linewidth}
      \centering
        \includegraphics[width=\linewidth]{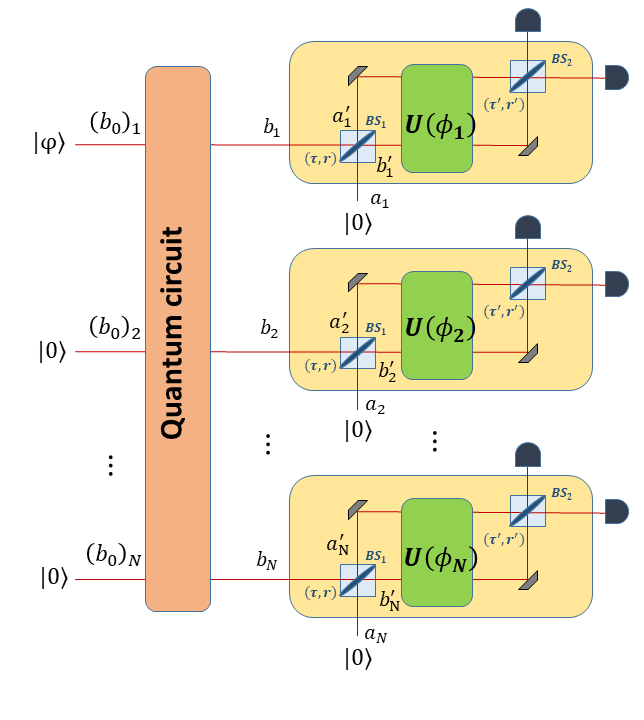}
(a) Entangled strategy  
     \label{fig:a}
    \end{minipage}%
   \begin{minipage}[b]{0.6\linewidth}
        \centering
        \includegraphics[width=\linewidth]{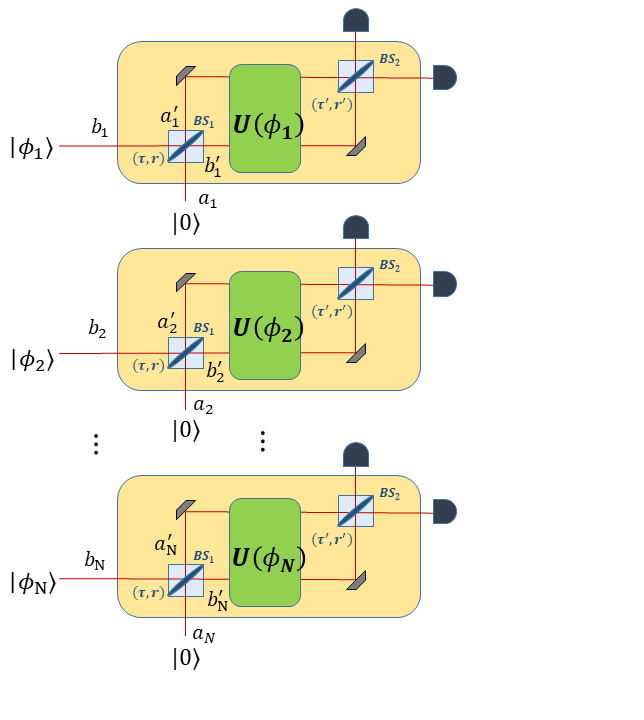}
       \label{fig:b}
(b) Separable strategy
    \end{minipage}%
    
     \caption{ Distributed quantum sensing setup using $N$ Mach–Zehnder (MZI) interferometers. 
(a) Entangled strategy, where a state $|\varphi\rangle$, distributed by an $N$-mode quantum circuit, feeds the various MZIs in an entangled form. 
(b) A separable strategy in which each MZI independently receives a state $|\varphi_k\rangle$. 
In both cases, $\phi_1, \dots, \phi_N$ represent the relative phase shifts introduced in each interferometer.}
    \label{model}

\end{figure}

The proposed distributed quantum sensing protocol studied in this work is based on a network of $N$ Mach--Zehnder interferometers (MZI). Each Mach--Zehnder interferometer consists of two beam splitters, $\mathrm{BS}_1$ and $\mathrm{BS}_2$, characterized by transmission (reflection) coefficients $\tau$ ($r$) and $\tau'$ ($r'$), respectively, and two arms encoding the phase shift, as illustrated in Figure~\ref{model}.

The entire system under study is structured as follows: The first part concerns the initial quantum state preparation, which constitutes the input state of the MZIs network. This preparation can be carried out using two strategies: the entangled strategy, illustrated in Figure~\ref{model}(a), in which a quantum circuit (QC) creates correlations only between the modes \(b_{k}\), and the separable strategy, illustrated in Figure~\ref{model}(b), in which the different modes of the probe are independent. The initial state of the probe, before the first beam splitter $(BS_{1})$, is generally denoted \( |\psi_{\mathrm{in}}\rangle \), regardless of its correlation structure.
The second part corresponds to the probe-state evolution within the network of \(N\) Mach--Zehnder interferometers (MZI). The overall state after the first beam splitter is denoted by \( |\psi\rangle \), and the output state, after encoding the simultaneous phase shifts, is written as
\begin{equation}
|\psi_{\mathrm{out}}\rangle = \mathbf{U}(\boldsymbol{\phi})\,|\psi\rangle ,
\label{psiout}
\end{equation}
where
\begin{equation}
\mathbf{U}(\boldsymbol{\phi}) =  \bigotimes_{k=1}^N 
\exp\!\left[i\frac{\phi_{k}}{2}\left(\hat{N}_{\hat a'_k}-\hat{N}_{\hat b'_k}\right)\right],
\end{equation}
is the global unitary operator associated with the simultaneous encoding of the \(N\) unknown relative phase shifts 
\(\boldsymbol{\phi}=(\phi_1,\ldots,\phi_N)\). Here, \(\hat{N}_{l_k}=\hat{l}_k^{\dagger}\hat{l}_k\) (\( l=  a', b'\)) denotes the number operator corresponding to mode \(l\) in the \(k\)-th interferometer.

The state \( |\psi\rangle \) is obtained by applying the field operator transformations induced by the first beam splitter to \( |\psi_{\mathrm{in}}\rangle \):
\begin{equation}
\hat{a}'_k = \tau\, \hat{a}_k + r\, \hat{b}_k, 
\qquad 
\hat{b}'_k = r\, \hat{a}_k + \tau\, \hat{b}_k,
\label{eq:field_transformations}
\end{equation}
where the transmission coefficient \(\tau\) and reflection coefficient \(r\) fulfill the unitarity condition \(|\tau|^2 + |r|^2 = 1\) and \(\tau r^* = -\tau^* r\). Without loss of generality, we adopt the convention \(\tau^* r = i|\tau r|\).

 Finally, the third part corresponds to the detection of output modes. The choice of the measurement strategy constitutes an important aspect of
interferometric quantum metrology, and several detection schemes have been investigated in Mach--Zehnder interferometers, including photon-counting
measurements, parity detection, and intensity-difference detection ~\cite{BenAryeh2012,Huang2017,Liu2024}. In the present work, however,
this step is not explicitly considered. The achievable phase sensitivity may depend on both the quantum state employed at the interferometer inputs
and the measurement performed at the outputs. Therefore, we focus on the quantum Fisher information and the corresponding quantum Cram\'er-Rao bound as a measurement-independent characterization of the ultimate phase estimation precision.

\subsection{Quantum circuit action}

The two configurations illustrated in Figures \eqref{model}(a) and \eqref{model}(b) involve a total of $2N$ input modes. In this work, the $a_k$ modes are taken to be initially prepared in the vacuum state, whereas the $b_k$ modes receive a general quantum state generated by a quantum circuit (QC). The joint input state of the system is then written as
\begin{equation}
|\psi_{\mathrm{in}}\rangle = |\psi_{\mathrm{QC}}\rangle \otimes |0\rangle^{\otimes N},
\end{equation}
where the state $|\psi_{\mathrm{QC}}\rangle$ is prepared by the quantum circuit and given by
\begin {equation}
|\psi_{\mathrm{QC}}\rangle 
= \mathbf{U}_{\mathrm{QC}}
\bigotimes_{k=1}^{N} |\phi_k\rangle .
\end{equation}
In the entangled scheme, only the input mode $(b_0)_1$ is fed by a nontrivial state $|\phi\rangle$, while all other modes are initially in the vacuum state. i.e.
\begin{equation}
|\phi_k\rangle = |0\rangle \quad \text{for } k=2 \dots N .
\end{equation}
The resulting output state of the quantum circuit is
\begin{equation}
|\psi_{\mathrm{QC}}\rangle = \mathbf{U}_{\mathrm{QC}} \Big[ |\phi\rangle \otimes |0\rangle^{\otimes(N-1)} \Big].
\end{equation}

The general state $|\phi\rangle$ admits the following decomposition in the Fock basis
\begin{equation}
|\phi\rangle = \sum_{\eta=0}^{\infty} c(\eta)\,|\eta\rangle ,
\end{equation}
where $|\eta\rangle = (\hat{b}_{0}^\dagger)^\eta / \sqrt{\eta!}\,|0\rangle$. We use the properties $[ |0 \rangle \otimes |0 \rangle \otimes \dots\otimes |0 \rangle ]=  \mathbf{U}_{\mathrm{QC}} [ |0 \rangle \otimes |0 \rangle \otimes \dots\otimes |0 \rangle ]$ and $\mathbf{U}_{\mathrm{QC}}^{\dagger} \mathbf{U}_{\mathrm{QC}} =  \mathbb{I} $, we can obtain
\begin{equation}
 |\psi_{\mathrm{QC}}\rangle= \sum_{\eta=0}^{\infty} c(\eta) \frac{(\mathbf{U}_{\mathrm{QC}}\,(\hat b_{\mathrm{0}}^\dagger)_1\,\mathbf{U}_{\mathrm{QC}}^\dagger)^{\eta}}{\sqrt{\eta!}} [ |0 \rangle \otimes |0 \rangle \otimes \dots\otimes |0 \rangle ]. \label{PSI} 
\end{equation}
The quantum circuit induces a linear transformation of the creation operators.  In the Heisenberg representation, this transformation is written as 
\begin{equation}
\mathbf{U}_{\mathrm{QC}}\,(\hat b_{\mathrm{0}}^\dagger)_1\,\mathbf{U}_{\mathrm{QC}}^\dagger = \sum_{k=1}^{N} (\mathbf{U}_{\mathrm{QC}})_{k1}\,\hat b_k^\dagger .
\end{equation}

By limiting ourselves to transformations with real coefficients, the coefficients of the unitary matrix $ \mathbf{U}_{\mathrm{QC}}$ can be parameterized in the form
\begin{equation}
(\mathbf{U}_{\mathrm{QC}})_{k1} = \sqrt{P_k}, \qquad P_k \ge 0, \qquad \sum_{k=1}^{N} P_k = 1,
\end{equation}
which leads to the relation
\begin{equation}
\mathbf{U}_{\mathrm{QC}}\,
(\hat b_{\mathrm{0}}^\dagger)_1\,
\mathbf{U}_{\mathrm{QC}}^\dagger
= \sum_{k=1}^{N} \sqrt{P_k}\,\hat b_k^\dagger, \label{hsn}
\end{equation}

where $P_k$ denotes the probability of photon detection in output mode $k$ of QC, obtained from the expectation value of the number operator.
\begin{equation}
P_k =
\frac{\langle \psi_{\mathrm{QC}} | \hat b_k^\dagger \hat b_k | \psi_{\mathrm{QC}} \rangle}{\sum_{k=1}^{N}
\langle \psi_{\mathrm{QC}} | \hat b_k^\dagger \hat b_k | \psi_{\mathrm{QC}} \rangle },
\end{equation}

Under the action of the quantum circuit, the input state of the QC is redistributed coherently across all QC output modes. The output state of QC can then be written as
\begin{equation}
|\psi_{\mathrm{QC}}\rangle
= \sum_{\{\eta_k\}}
c_{\mathrm{QC}}(\eta_1,\ldots,\eta_N)
\bigotimes_{k=1}^{N} |\eta_k\rangle .
\label{QCme}
\end{equation}
By substituting Eq.~\eqref{hsn} into the QC state \eqref{PSI} and applying the multinomial theorem, the resulting generalized multinomial amplitudes are given by
\begin{equation}
c_{\mathrm{QC}}(\eta_1,\ldots,\eta_N)
=
\sqrt{\frac{\eta!}{\eta_1!\cdots\eta_N!}}
\prod_{k=1}^{N} P_k^{\eta_k/2}\,
c(\eta),
\:
\eta=\sum_{k=1}^{N} \eta_k .
\end{equation}
This result shows that the quantum circuit coherently redistributes the input state over the $N$ output modes according to a multinomial structure characterized by the probabilities $\{P_k\}$. This structure constitutes the natural starting point for the analysis of the entangled strategy developed in the rest of this work.\\
In the \emph{separable} scheme, the quantum circuit is trivial, $\mathbf{U}_{\mathrm{QC}} = \mathbb{I}$, so that each mode $b_k$ receives an independent state $|\phi_k\rangle$. The global state injected into the sensor array, consisting of $N$ Mach--Zehnder interferometers, is then written as
 \begin {equation}
|\psi_{\mathrm{QC}}\rangle 
= \bigotimes_{k=1}^{N} |\phi_k\rangle . \label{QCsp}
\end{equation}
\subsection{Quantum Cramér-Rao bound sensitivity} 
In this work, we first consider the ideal pure and lossless configuration. The QFIM is calculated first, as it is the fundamental tool for evaluating the ultimate precision achievable by the network.  More specifically, the QFI quantifies the intrinsic sensitivity of a network of Mach--Zehnder interferometers (MZI)  to infinitesimal variations in the phases to be estimated. From the QFIM, we derive the  QCRB, providing a fundamental lower limit on the variance of any unbiased estimator.  thus providing a rigorous framework for comparing the performance of the two strategies studied.  It is essential to note that the second beam splitter (BS2) does not impact the QFI, as it corresponds to a unitary transformation that is independent of the phase parameters.
Using Eqs.~\eqref{FQ} and \eqref{psiout},the quantum Fisher information matrix elements, denoted $( \mathcal{F}_{Q})_{kl}$, are given by
\begin{align}
   ( \mathcal{F}_{Q})_{kl}  =  (\mathcal{\tau}^2-r^2)^2( & \langle  \psi_{QC} \vert \hat{n}_k \hat{n}_l  \vert \psi_{QC} \rangle  -\langle \psi_{QC} \vert \hat{n}_k  \vert \psi_{QC} \rangle\langle \psi_{QC} \vert \hat{n}_l \vert \psi_{QC} \rangle)  \nonumber  \\  +   & 4|\tau r|^2 \langle \psi_{QC} \vert \hat{n}_k  \vert \psi_{QC} \rangle \delta_{kl}, 
   \label{Q1}
\end{align}
where $\hat{n}_k= \hat{b}_k^{\dagger} \hat{b}_k$.
\subsubsection{Mode-entangled strategy}

Starting from equation \eqref{QCme} and using the multinomial distribution properties\cite{johnson1960approximation}, we perform the following calculation:
\begin{align}
 &\langle \psi_{QC} \vert \hat{n}_k  \vert \psi_{QC} \rangle = \langle \phi \vert \hat n_{0} \vert \phi \rangle \:P_{k}, \\
   & \langle  \psi_{QC} \vert \hat{n}_k \hat{n}_l  \vert \psi_{QC} \rangle  =   \left( \langle \phi \vert \hat n_{0}^2 \vert \phi \rangle -\langle \phi \vert \hat n_{0} \vert \phi \rangle \right)\:P_{k}  P_{l} +\langle \phi \vert \hat n_{0} \vert \phi \rangle \; P_{k}\; \delta_{kl}.
\end{align}qua
Where $\hat{n}_{0}= (\hat{b}_{0}^\dagger)_{1} (\hat{b}_{0})_{1} $. Then, the components of the QFIM are expressed as follows:
\begin{align}
    (\mathcal{F}_{Q})_{kl}  = & (\mathcal{|\tau|}^2-|r|^2)^2 (\Delta^2( \hat n_{0})_{\vert{\phi}\rangle}-  \langle \phi \vert  \hat n_{0}  \vert \phi \rangle) \:P_{k}  P_{l}  \\ \nonumber   & +\langle \phi \vert  \hat n_{0}  \vert \phi \rangle\; P_{k} \;\delta_{kl}.
    \label{Q1me}
\end{align}
 We note that the entanglement between modes, induced by the action of the quantum circuit $(QC)$, leads to the appearance of off-diagonal elements in the quantum Fisher information matrix. Conversely, for the separable strategy, the matrix remains strictly diagonal.
 To highlight this, let us first note that the quantum Fisher information matrix can be rewritten as
\begin{equation}
\mathcal{F}_{Q}
= \chi\, \mathbf{g}\mathbf{g}^{T} + \mathcal{D},
\end{equation}
in which the coefficient $\chi$ is given by
\begin{equation}
\chi
= \bigl(|\tau|^{2}-|r|^{2}\bigr)^{2}
\left(
\Delta^{2}( \hat n_{0})_{\vert \phi\rangle}
- \langle \phi \vert  \hat n_{0} \vert \phi \rangle
\right).
\end {equation}
The vector $\mathbf{g} = (P_{1}, \ldots, P_{N})^{T}$ denotes a real vector of dimension $N$.
The matrix $\mathcal{D}$ is diagonal of size $N \times N$, and its diagonal elements are given by
\begin{equation}
(\mathcal{D})_{kk}
= 
\langle \phi \vert \hat{n} \vert \phi \rangle\, P_{k}.
\end{equation}
The analytical inverse of the quantum Fisher information matrix is derived using the Sherman–Morrison formula, giving:
 \begin{equation}
    \mathcal{F}_{Q}^{-1} = \mathcal{D}^{-1} - \frac{ \chi}{1+ \chi\mathbf{g}^{T} \mathcal{D}^{-1} \mathbf{g} }(\mathcal{D}^{-1} \mathbf{g})(\mathcal{D}^{-1} \mathbf{g})^{T}.
 \end{equation}
\\
Substituting this last expression into equation \eqref{QCRB}, we derive the quantum Cramér-Rao bound for the entangled strategy.
\begin{align}
\Delta^2(\mathbf{\mu} \cdot \boldsymbol{\phi})_{QCRB}^{me}
&= 
\sum_{k=1}^{N} \frac{\mu_k^2}{\langle \phi \vert  \hat n_{0} \vert \phi \rangle\, P_k} 
-
\frac{  \, \chi}{\langle \phi \vert  \hat n_{0} \vert \phi \rangle (\langle \phi \vert  \hat n_{0} \vert \phi \rangle + \chi )}
\left(\sum_{k=1}^{N} \mu_k \right)^2.
\label{QCRBe1}
\end{align}
 The QCRB derived above is asymptotically saturable due to the commutativity of the generators, i.e., $[H_j, H_k] = 0$, where $H_k = \frac{1}{2}(\hat{N}_{a'_k} - \hat{N}_{b'_k})$, and the corresponding optimal precision can be attained via local photon-number measurements at the output of each interferometer \cite{Pezze2017}.
 
\subsubsection{Mode-separable strategy}
For input state \eqref{QCsp}, the equation \eqref{Q1} becomes:
\begin{equation}
    ( \mathcal{F}_{Q})_{kl}  = (\mathcal{\tau}^2-r^2)^2 \Delta^2(\hat{n}_k)_{\vert{\phi_k}\rangle} +4|\tau r|^2 \langle \phi_{k} \vert \hat{n}_k  \vert \phi_{k} \rangle  \label{Q1ms} .
\end{equation}
Based on equation \eqref{QCRBs}, the Cramér–Rao quantum bound is written as:
\begin{equation}
 \Delta^2(\mathbf{\mu} \cdot \boldsymbol{\phi})_{Q}^{ms} = \sum_{k=1}^{N} \frac{\mu_{k}^{2}}{ (\mathcal{\tau}^2-r^2)^2 \Delta^2(\hat{n}_k)_{\vert{\phi_k}\rangle} +4|\tau r|^2 \langle \phi_{k} \vert \hat{n}_k  \vert \phi_{k} \rangle }.
 \label{QCRBs1}
\end{equation}

Until now, the study of the network has been conducted in general terms, without specifying a particular input state for the quantum circuit. We now consider the Perelomov coherent states corresponding to the non-compact group SU(1,1), denoted $|\xi\rangle$, as the input state of the QC. In the entangled strategy, the quantum circuit is fed by the state $\vert \varphi\rangle = |\xi\rangle$, while in the separable strategy, each Mach--Zehnder interferometer receives the state $\vert \varphi\rangle_k = |\xi\rangle_k$.   
These states are known to exhibit intrinsic squeezing properties arising from the non-compact SU(1,1) algebra, which makes them highly suitable for improving phase sensitivity in interferometric and distributed quantum sensing networks. The squeezing and statistical properties of SU(1,1) Perelomov coherent states have been extensively studied in the literature, showing that they can reduce quantum fluctuations below the standard quantum limit and serve as valuable resources for quantum metrology\cite{Buzek1989}. The average photon number and the corresponding variance for these states are given by:
\begin{equation}
\overline{n}_{\xi} = a (\cosh v - 1), \qquad
\Delta^{2}(\hat{n})_{\vert \xi \rangle} = \frac{a}{2} \sinh^{2}(v),
\end{equation}
where $v$ is the squeezing parameter. 

 
The SU(1,1) Perelomov coherent state can be understood directly from its photon-number statistics.  Unlike an ordinary coherent state, whose photon-number variance is  equal to itsaverage photon number, the SU(1,1) Perelomov coherent state  exhibits a non-Poissonian photon-number distribution characterized by
\begin{equation}
\Delta^{2}(\hat{n})_{\vert \xi \rangle} = \overline{n}_{\xi} + \frac{\overline{n}_{\xi}^{2}}{2a}. \label{sqz}
\end{equation}

The second term, $\bar n_\xi^2/(2a)$, represents an additional quadratic contribution to the photon-number fluctuations. This term becomes increasingly important as the average photon number increases and provides the nonclassical resource exploited by the proposed sensing scheme. Therefore, the  SU(1,1) Perelomov coherent state is particularly suitable for the present distributed sensing architecture because it combines nonclassical photon-number statistics with the ability to be coherently redistributed among multiple sensor modes by the quantum circuit.
Substituting this expression \eqref{sqz} in equations \eqref{QCRBe1} and \eqref{QCRBs1}, we can directly obtain the QCRBs associated with our QC input state for the two strategies;

\begin{align}
 \Delta^2(\mathbf{\mu} \cdot \boldsymbol{\phi})_{QCRB}^{me}
 = \sum_{k=1}^{N} \frac{\mu_k^2}{\overline{n}_{\xi}\; P_k} -\frac{  \, \frac{\overline{n}_{\xi}^2}{2a} \;(|\mathcal{\tau}|-|r|^2)^2\; \left(\sum_{k=1}^{N} \mu_k \right)^2  }{(\overline{n}_{\xi}^2 + \frac{\overline{n}_{\xi}^3}{2a_k} \; (|\mathcal{\tau}|^2-|r|^2)^2 )}   
\label{QCRBme1}
\end{align}
\begin{equation}
 \Delta^2(\mathbf{\mu} \cdot \boldsymbol{\phi})_{QCRB}^{ms} = \sum_{k=1}^{N} \frac{\mu_{k}^{2}}{ \overline{n}_{\xi_{k}}+ \frac{\overline{n}_{\xi_{k}}^2}{2a_k} (\mathcal{\tau}^2-r^2)^2 },
 \label{QCRBms1}
\end{equation}
    
\section{Optimized quantum Cramér-Rao bound sensitivity}
In the present section, we first optimize the network of $N$ Mach-Zehnder interferometers by minimizing the quantum Cramér-Rao bounds (QCRB) for each of the two strategies considered. We then compare these two strategies based on the optimal QCRB values obtained.  In the optimization procedure for a Mach-Zehnder interferometer network, the parameter $a$ characterizing the Perelomov coherent state SU(1,1) is assumed to be identical for the entire network and thus remains independent of the mode index $k$. As a consequence, the average photon number $\overline{n}_{\xi}$ is governed only by the squeezing parameters $\xi_k$, which determine the photon distribution within the network. 
\subsection{Entangled strategy}
 We begin by optimizing the entangled strategy. To this end, equation \eqref{QCRBme1} is optimized with respect to the coefficients \(P_k\), which characterize the transformation of the quantum circuit \(U_{\mathrm{QC}}\).
\begin{equation}
    \Delta^2(\boldsymbol{\mu} \cdot \boldsymbol{\phi})_{Q}^{\rm me} =\min\limits_{P_{1}, \dots, P_{N}}
 \Delta^2(\boldsymbol{\mu} \cdot \boldsymbol{\phi})_{QCRB}^{\rm me}.
 \label{minQme}
\end{equation}
 The minimization of the QCRB for the entangled strategy is performed using the Lagrange multiplier method. This approach allows the normalization constraint on the coefficients \(\sum_{k=1}^{N} P_k = 1\) to be incorporated and leads to the optimal Cramér-Rao bounds for multiparametric estimation. The associated Lagrangian, constructed from the dominant term \(\sum_{k=1}^{N} \mu_k^2/(\overline{n}_\xi\, P_k)\), is written as
\begin{equation}
\mathcal{L} = \sum_{k=1}^{N} \frac{\mu_k^2}{\overline{n}_\xi\, P_k} + \lambda \left( \sum_{k=1}^{N} P_k - 1 \right),
\end{equation}
where \(\lambda\) is the Lagrange multiplier. By differentiating $\mathcal{L}$ with respect to $P_k$ and imposing $\partial \mathcal{L}/\partial P_k = 0$, we  find the optimal condition for the coefficients:
\begin {equation}
P_l = \frac{|\mu_l|}{\sum_{k=1}^{N} |\mu_k|}. \label{popt}
\end{equation}
This equation shows that the optimal distribution of quantum resources across the network depends directly on the weighting vector $\boldsymbol{\mu}$. In particular, the sensors associated with the largest components of $\boldsymbol{\mu}$ receive a larger fraction of photons. This means, on average, that quantum resources are concentrated in the sensors most relevant to the estimation of the parameter in focus. Substituting this solution \eqref{popt} into the expression \eqref{QCRBme1}, the optimal QCRB is given by
\begin{equation}
\Delta^2(\boldsymbol{\mu}\cdot\boldsymbol{\phi})_{Q}^{\mathrm{me}}
=
\frac{\left(\sum_{k=1}^{N} |\mu_k|\right)^2}
{\overline{n}_{\xi} 
+ \dfrac{\overline{n}_{\xi}^2}{2a}\left(|\tau|^2-|r|^2\right)^2}.
\label{Qme}
\end{equation}
This expression corresponds to the maximum precision achievable by the entangled strategy for an optimal choice of quantum circuit parameters.
equation \eqref{Qme} can be bounded by two specific limits depending on the structure of vector \(\boldsymbol{\mu}\), which influences the linear combination to be estimated\cite{Malitesta2023}. The upper bound is obtained when all modes contribute symmetrically,
\[
\mu_k = \frac{1}{N}, \quad k = 1, \dots, N,
\]
which correspond to the estimation of the average phase 
\(\phi_{\rm avg} = (\phi_1 + \cdots + \phi_N)/N\). In this case, resource exploitation is entirely symmetric, and the optimal quantum sensitivity for the entangled strategy is written as:
\begin{equation}
\Delta^2(\boldsymbol{\mu_{\rm av}} \cdot \boldsymbol{\phi})_{Q}^{me} =
\frac{1}{\overline{n}_\xi + \frac{\overline{n}_\xi^2}{2a} \left(|\tau|^2 - |r|^2\right)^2}.
\label{Qmeav}
\end{equation}

\begin{figure}[htbp]
     
      \centering
        \includegraphics[width=\linewidth]{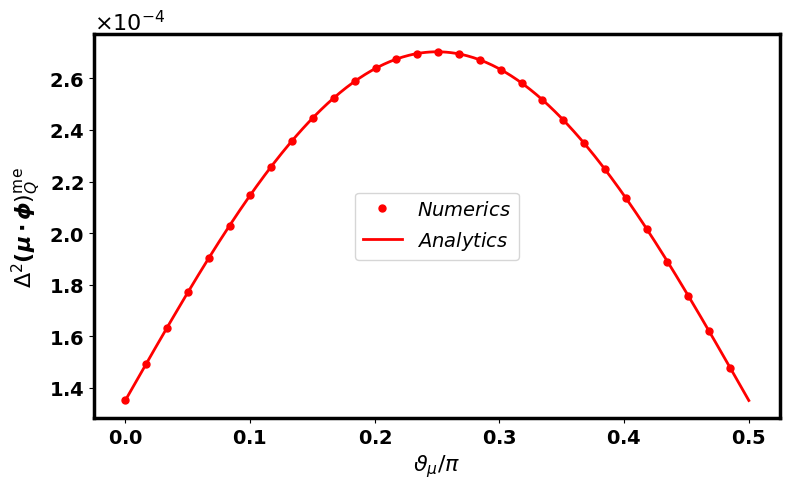}
     \caption{ The optimal sensitivity for the entangled strategy as a function of $\boldsymbol{\mu}$ for $N=2$, $\overline{n}_{\xi}=100$, $a=0{.5}$ and $\tau=0{.8}$. The red dots represent the numerical evaluation of equation \eqref{minQme} and the red line represents the analytical prediction defined by \eqref{Qme}. The vector  $\boldsymbol{\mu}$ is given by  
$
\boldsymbol{\mu}^{T}=\frac{1}{\sqrt{2}}\left(\cos\vartheta_\mu,\,\sin\vartheta_\mu\right),\:
\vartheta_\mu\in\left[0,\frac{\pi}{2}\right].
$}  
      \label{mu}   
\end{figure}

\begin{figure}[htbp]
   \centering
       \begin{minipage}[b]{1\linewidth}
      \centering
      \begin{overpic}[width=\linewidth]{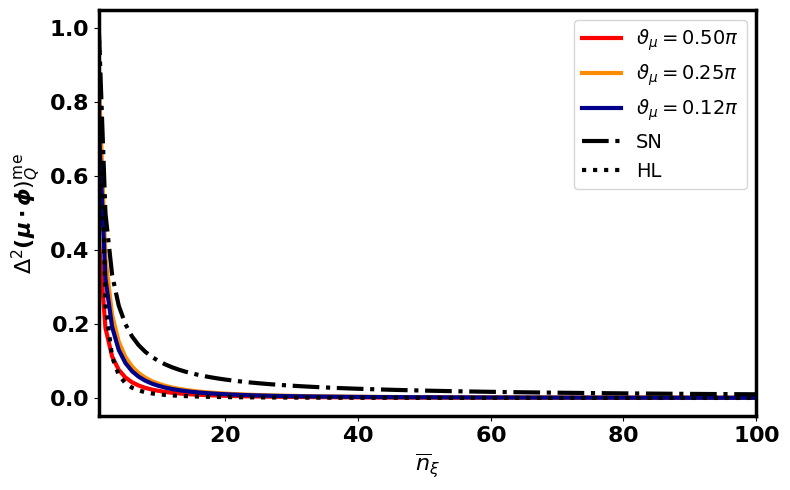}
         \put(18,19){\includegraphics[width=0.55\linewidth]{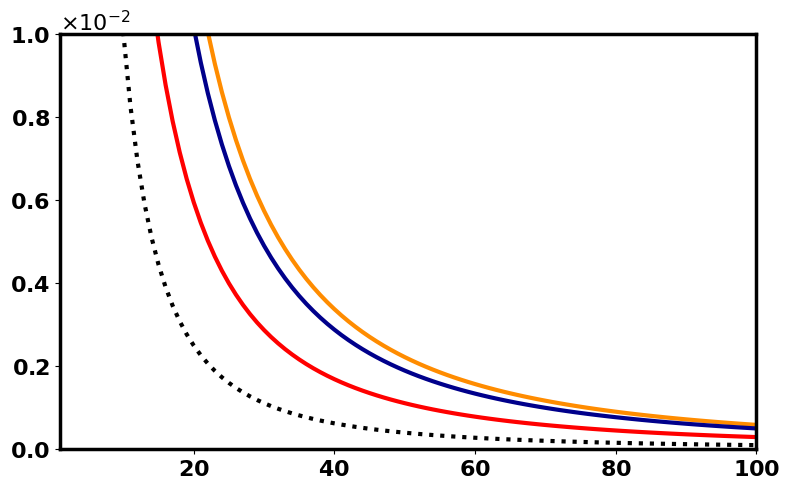}}
      \end{overpic}
(a)
		\end{minipage}\\
        \begin{minipage}[b]{1\linewidth}
			\centering
			\includegraphics[width=\linewidth]{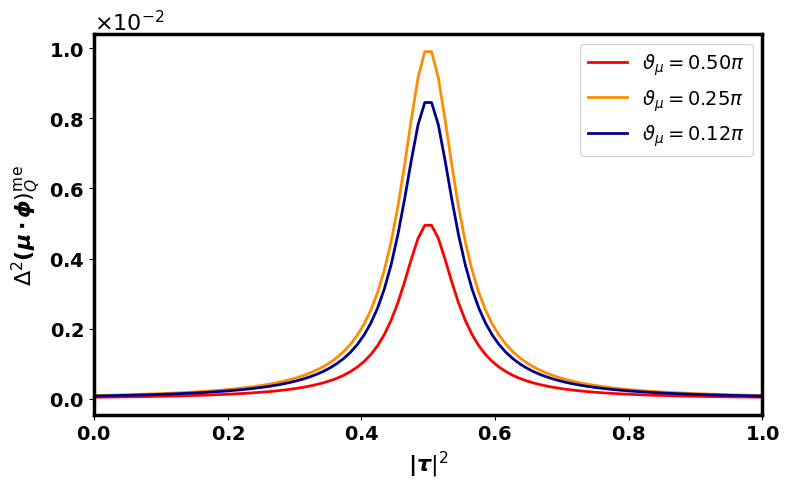}
  (b)
		\end{minipage}%
		\caption {Optimal quantum sensitivity of the entanglement strategy \eqref{Qme} ($N=2$, $a=0.5$) for different values of $\boldsymbol{\mu}$: (a) as a function of $\overline{n}_{\xi}$ ($|\tau|^2 = 0.7$); (b) as a function of $|\tau|^2$ ($\overline{n}_{\xi} = 100$). The same vector $\boldsymbol{\mu}$ as the Fig \eqref{mu}.}
        \label{me2d}
 \end{figure}

The lower bound, on the other hand, corresponds to the estimation of a single parameter, with 
\[
|\mu_l| = \frac{1}{\sqrt{N}}, \quad |\mu_k| = 0 \quad \text{for } k \neq l,
\]
which satisfies the normalization condition $|\mathbf \mu|^2=1/N$ and reduces the problem to a single-phase estimation, formally equivalent to that performed in a single Mach--Zehnder interferometer. The optimal QCRB then becomes
\begin{equation}
\Delta^2(\boldsymbol{\mu_{\rm sng}} \cdot \boldsymbol{\phi})_{Q}^{me} =
\frac{1}{N \left( \,\overline{n}_\xi +  \frac{\overline{n}_\xi^2 }{2a}\left(|\tau|^2 - |r|^2\right)^2 \right)}.
\label{Qmesng}
\end{equation}
Equation~\eqref{Qme} reveals that the estimation accuracy depends directly on the average photon number $\overline{n}_{\xi}$, the weight vector $\boldsymbol{\mu}$, and the interferometer configuration. For a balanced interferometer at $50:50$, the minimum variance scales as
\begin{equation}
\Delta^2(\boldsymbol{\mu}\cdot\boldsymbol{\phi})_{\mathrm{QCRB}}^{\mathrm{me}}
\sim
\frac{1}{\overline{n}_{\xi}},
\end{equation}
corresponding to the shot-noise scaling. This behavior changes as the first beam splitter is moved away from the balanced configuration. In contrast, optimizing the transmission coefficient of the first beam splitter of each MZI shows that the minimum QCRB is obtained at the extreme values $|\tau|^2=0$ or $|\tau|^2=1$, using the same Lagrange multiplier method introduced below. At these optimal transmission values and in the high-photon-number regime, the minimum QCRB behaves as
\begin{equation}
\Delta^2(\boldsymbol{\mu}\cdot\boldsymbol{\phi})_{\mathrm{QCRB}}^{\mathrm{me}}
\sim
\frac{1}{\overline{n}_{\xi}^{\,2}},
\end{equation}
corresponding to Heisenberg scaling. This result is consistent with and generalizes observations reported for single interferometers \cite{Ataman2022, Preda2019, Ataman2020}, which show that, for certain Gaussian and non-Gaussian input states and under specific constraints on the input parameters, beam-splitter imbalance can play an important role in phase-shift estimation.
The occurrence of the optimal sensitivity in the highly unbalanced regime may appear counterintuitive, since the first beam splitter
approaches a nearly fully transmissive or fully reflective configuration as $|\tau|^2$ approaches $0$ or $1$. The physical
origin of this behavior is related to the photon-number difference that generates the relative phase, $\hat{\mathcal{H}}_k=(\hat{N}_{a'_k}-\hat{N}_{b'_k})/2$. The first beam splitter determines how the photon-number fluctuations of the non-vacuum input contribute to this number difference. In the highly unbalanced limit, the splitting becomes strongly asymmetric, so that these fluctuations contribute with their largest weight to the variance of the phase generator. This effect is particularly important for the SU(1,1) Perelomov coherent state, whose photon-number variance contains the non-Poissonian quadratic term
$\overline{n}_{\xi}^{\,2}/(2a)$. The QFIM explicitly reflects this behavior through the factor $(|\tau|^2-|r|^2)^2$ multiplying this
quadratic contribution. As $|\tau|^2$ approaches $0$ or $1$, this factor approaches unity, showing that the photon-number fluctuations responsible for the enhanced sensitivity contribute maximally to the variance of the phase generator. Thus, although the first beam splitter approaches an extreme configuration, the phase sensitivity is not suppressed within the considered encoding scheme; the non-Poissonian fluctuations of the probe instead give their maximum contribution to the QFI, accounting for the minimum QCRB obtained in this regime.
\\
These analytical properties are confirmed by figure \eqref{mu} and figure \eqref{me2d} for $N=2$ and $a=0.5$.  In Fig.~\eqref{mu}, we compare the numerical evaluation of equation \eqref{minQme} (red dot) with the analytical prediction given by \eqref{Qme} (red line), as a function of $\boldsymbol{\mu}$ for $\overline{n}_{\xi}=100$ and   $\vert\tau \vert^2=0.8$. The excellent agreement observed over the entire interval $\vartheta_{\mu}\in[0,\pi/2]$ confirms the consistency between the direct evaluation of \eqref{minQme} and the analytical expression in \eqref{Qme}.
 The sensitivity depends heavily on the vector $\boldsymbol{\mu}$: it reaches its minimum for $\vartheta_\mu\simeq 0$ and $\vartheta_\mu\simeq \pi/2$,  which corresponds to the single-phase estimation \eqref{Qmesng}, while its maximum is obtained in the case of the average phase estimation \eqref{Qmeav}, i.e. $\vartheta_\mu\simeq \pi/4$. The sensitivity behavior also reproduces that reported in reference~\cite{Malitesta2023} for distributed Mach--Zehnder interferometers injected with coherent and squeezed vacuum states. The key distinction lies in the approach adopted to quantify sensitivity. Figure~\eqref{me2d} illustrates the optimized QCRB of the entangled strategy, given by Eq.~\eqref{Qme}, under different parameter regimes.
In panel (a), the QCRB is plotted as a function of $\overline{n}_{\xi}$ for $|\tau|^2=0.7$ and different values of $\vartheta_{\mu}$, together with the shot-noise scaling (black dash-dotted line) and the Heisenberg scaling (black dotted line). The figure shows a monotonic improvement in precision as the average photon number increases. The estimation precision exceeds the shot-noise scaling and approaches the Heisenberg scaling as $\overline{n}_{\xi}\gg1$. The inset provides a magnified view of the interval $[0,0.01]$, highlighting the variations in the low-QCRB region. 
In panel (b), the transmission coefficient of the first beam splitter is varied to investigate its influence on the estimation precision. The results are consistent with the analytical behavior discussed above, showing a progressive improvement in precision as the beam-splitter configuration becomes increasingly unbalanced. This behavior is consistently observed for all the considered weight vectors $\boldsymbol{\mu}$, indicating that the influence of the first beam-splitter imbalance on the estimation precision is preserved for different linear combinations of the phase parameters.
 
     \subsection{Separable strategy}
 Although our focus is primarily on the entangled strategy, we also evaluate the minimum of the separable strategy sensitivity for comparison purposes.
 To this end, equation \eqref{QCRBms1} is performed based on the average photon number associated with the \( N \) $SU(1,1)$ Perelomov coherent states. 
\begin{equation}
    \Delta^2(\boldsymbol{\mu} \cdot \boldsymbol{\phi})_{Q}^{\rm ms} =\min\limits_{\overline{n}_{\xi_1}, \dots, \overline{n}_{\xi_N}}
 \Delta^2(\boldsymbol{\mu} \cdot \boldsymbol{\phi})_{QCRB}^{\rm ms}.
 \label{minQms}
\end{equation}
For this strategy, analytical minimization of the  QCRB is generally difficult to achieve. Nevertheless, it is possible to define analytical lower and upper bounds as in the entangled strategy, namely
\begin{equation}
\Delta^2(\boldsymbol{\mu_{sng}}\!\cdot\!\boldsymbol{\phi})_{Q}^{ms} 
\leq \Delta^2(\boldsymbol{\mu}\!\cdot\!\boldsymbol{\phi})_{Q}^{ms} 
\leq \Delta^2(\boldsymbol{\mu_{av}}\!\cdot\!\boldsymbol{\phi})_{Q}^{ms},
\label{msen}
\end{equation}
with
\begin {equation}
\Delta^2(\boldsymbol{\mu_{sng}}\!\cdot\!\boldsymbol{\phi})_{Q}^{ms}
=
\frac{1}
{N\left(\overline{n}_{T}
+
\frac{\overline{n}_{T}^{2}}{2a}(|\mathcal{\tau}|^2-|r|^2)^2\right)},
\label{mssng}
\end{equation}
and
\begin{equation}
\Delta^2(\boldsymbol{\mu_{av}}\!\cdot\!\boldsymbol{\phi})_{Q}^{ms}
=
\frac{1}
{\overline{n}_{T}
+
\frac{\overline{n}_{T}^{2}}{2Na}(|\mathcal{\tau}|^2-|r|^2)^2}.
\label{msav}
\end{equation}
The proof of this inequality and expressions~\eqref{mssng} and~\eqref{msav} is provided in the Appendix.
\begin{figure}[htbp]
 
     \begin{minipage}[b]{1\linewidth}
      \centering
        \includegraphics[width=\linewidth]{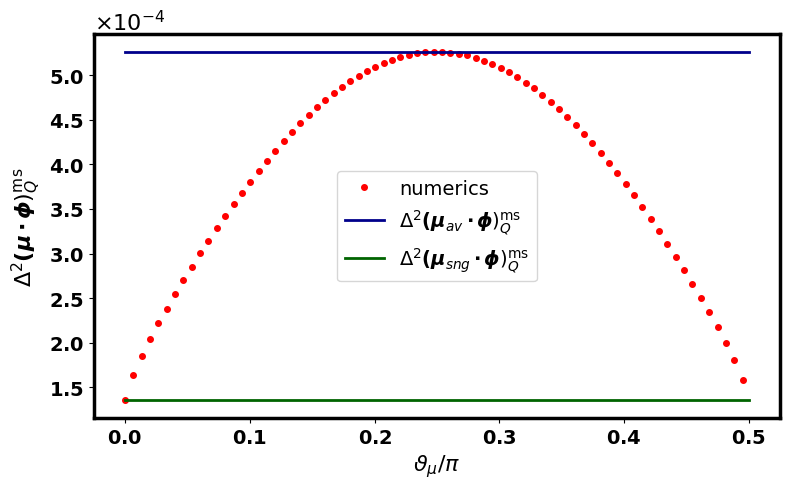}
(a) \label{a}
    \end{minipage} \\
    \begin{minipage}[b]{1\linewidth}
        \centering
        \includegraphics[width=\linewidth]{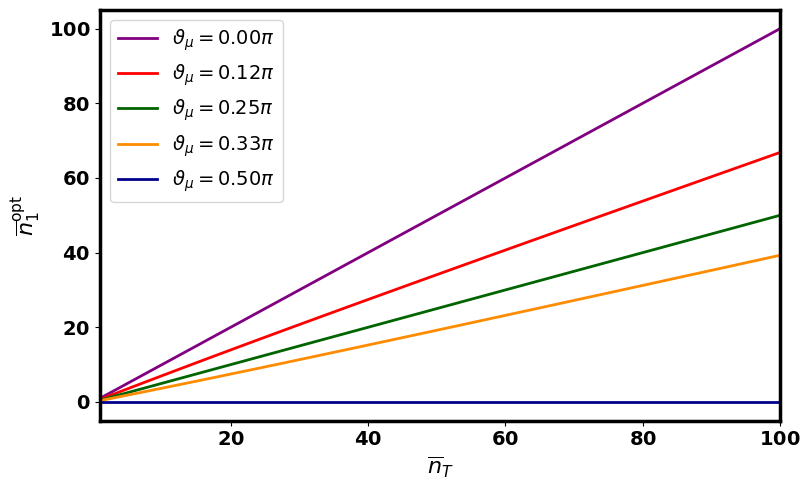}
        \label{b}
(b)
    \end{minipage}
     \caption{(a) Optimal quantum sensitivity obtained by numerically optimizing the quantum Cramér–Rao bound (QCRB) under the separable strategy \eqref{minQms} (red points), as a function of $\boldsymbol{\mu}$. The reference curves correspond to \eqref{msav} (blue line) and \eqref{mssng} (green line). 
(b) Optimal average number of photons in the first mode as a function of the total average number of photons, for different values of $\boldsymbol{\mu}$. 
The parameters are those from Figure \eqref{mu}.}  
     \label{ms2}     
\end{figure}

In this context, a numerical optimization of~\eqref{minQms} is plotted in Fig. \eqref{ms2} for $N=2$, $a=0{.5}$ and $\tau=0{.8}$.
 Panel (a) illustrates the optimal quantum sensitivity $\Delta^2(\boldsymbol{\mu} \cdot \boldsymbol{\phi})_{Q}^{\rm ms}$  for the separable strategy as a function of $\boldsymbol{\mu}$  (we use the same vector $\boldsymbol{\mu}$ as the one used in the entangled strategy).  
The red dots correspond to the results obtained by numerical optimization of the  QCRB \eqref{minQms}, while the horizontal lines represent the analytical results: the blue line describes the optimal average phase sensitivity \eqref{msav}, and the green line corresponds to that of a single-phase estimation \eqref{mssng}.
The sensitivity shows a marked dependence on $\mu$ and a symmetric behavior similar to that observed for the entangled strategy, with a maximum reached around $\vartheta_\mu\simeq \pi/4$, where the numerical optimization coincides with the analytical limit \eqref{msav} (blue line). Conversely, at the extremes  $\vartheta_\mu\simeq 0$ and $\vartheta_\mu\simeq \pi/2$, corresponding to configurations dominated by a single component of $\boldsymbol{\mu}$, the numerical results exactly overlap with the single-phase limit \eqref{mssng} (green line). 
Overall, the numerically optimized sensitivity remains strictly confined between the two analytical limits over the entire interval of $\vartheta_\mu$, confirming the validity of the inequality given in \eqref{msen}.
Figure (b) shows the optimal average photon number  $\overline{n}_1^{\mathrm{opt}}$, obtained by numerical optimization, in terms of the total average photon number $\overline{n}_T$, for $d=2$ and various values of the parameter $\vartheta_{\mu}$. We observe that, for a given value of $\vartheta_{\mu}$, the optimal photon distribution increases linearly with $\overline{n}_T$, indicating that an increase in photonic resources results in a proportional increase in the optimal allocation. The parameter $\vartheta_{\mu}$ only modifies the slope of this linear dependence and thus determines the fraction of photons allocated to each mode, reflecting how the photonic resources must be distributed between the two modes with the aim of optimizing the sensitivity of the phase estimation.
\subsection{Comparison of optimal entangled and separable strategies}
\begin{figure}
    \centering
    \includegraphics[width=0.9\linewidth]{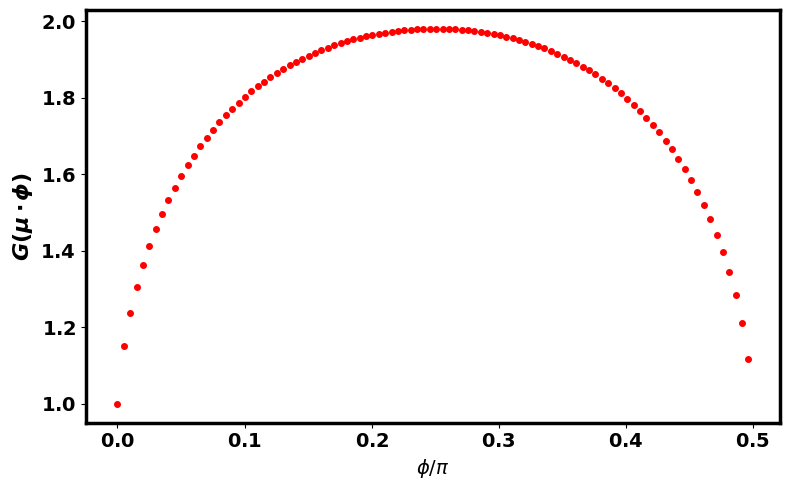}
    \caption{The numerical evaluation of the sensitivity gain given by Eq. \eqref{gn} as a function of $\boldsymbol{\mu}$. The other parameters are identical to that of Fig. \eqref{mu}}
    \label{gain}
\end{figure}

In this section, we analyze the gain factor \( G(\boldsymbol{\mu} . \boldsymbol{\phi})\), introduced as the ratio between the optimal QCRB associated with separable and entangled strategies, assuming that both strategies are fed by the same average total photon number, i.e.  $\overline{n}_{\xi} = \overline{n}_T$. 
 This analysis allows us to systematically quantify the advantage of mode entanglement for phase-shift estimation accuracy by comparing the performance of the two strategies. This gain is expressed as
 
 \begin{equation}
   G(\boldsymbol{\mu} . \boldsymbol{\phi}) = \frac{\Delta^2(\boldsymbol{\mu} \cdot \boldsymbol{\phi})_{Q}^{\rm ms}}{\Delta^2(\boldsymbol{\mu} \cdot \boldsymbol{\phi})_{Q}^{\rm me}}. \label{gn}
 \end{equation} 
In the case where $\mu=\mu_{\mathrm{sng}}$, the gain takes the value
\begin{equation}
G(\mu_{\mathrm{sng}}\!\cdot\!\phi) = 1.
\end{equation}
On the other hand, for $\mu=\mu_{\mathrm{av}}$, the gain factor is given by.
\begin{equation}
     G(\mu_{av}\cdot \phi) =\dfrac{2a\, + \;\overline{n}_{T}(\tau^2-r^2)^2}{2a \,
+  \dfrac{\overline{n}_{T}}{N}(\tau^2-r^2)^2} \label{gav}
\end{equation}
We note that $G(\mu_{av}\cdot \phi)=1$ for a balanced interferometer. On the other hand, an imbalance of the beam splitter, as well as increasing the number of interferometers $N$ or the total average photon number $\overline{n}_{T}$, yields to an improvement in the gain factor. Fig.~\eqref{gain} shows, based on a numerical evaluation, the gain~\eqref{gn} as a function of  $\boldsymbol{\mu}$. We observe that the gain reaches its maximum, close to $N$,  in the average phase estimation regime.  In this configuration, all MZIs contribute in a balanced manner, allowing the entangled strategy to fully exploit quantum correlations and reveal its truly collective nature. 
In contrast, in the single-phase estimation case, where a single MZI is used ($\vartheta_\mu \simeq 0$ or $\vartheta_\mu \simeq \pi/2$),  the gain reduces to unity, since no correlations are utilized. This shows that the entangled strategy always outperforms the separable strategy, 
and that its advantage is maximal in the regime of average-phase estimation, thus highlighting the key role of mode entanglement in optimizing the protocol’s performance.

\begin{figure}[htbp]
     \begin{minipage}[b]{1\linewidth}
      \centering
        \includegraphics[width=\linewidth]{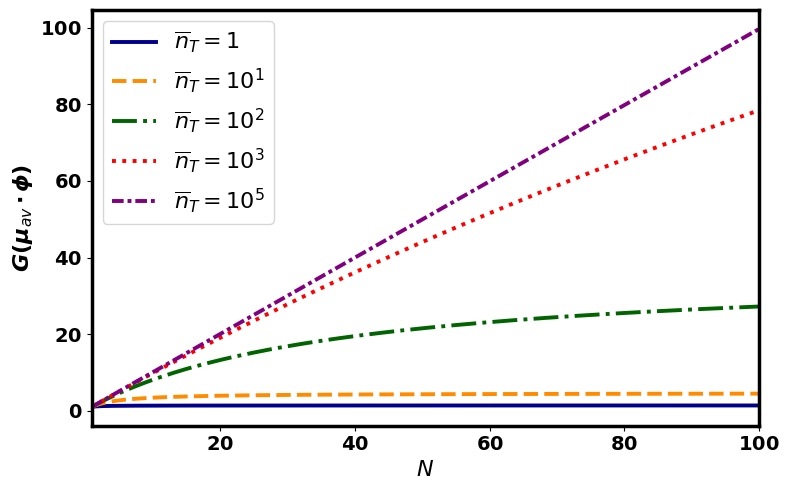}
(a) \label{a}
    \end{minipage}\\ 
    \begin{minipage}[b]{1\linewidth}
        \centering
        \includegraphics[width=\linewidth]{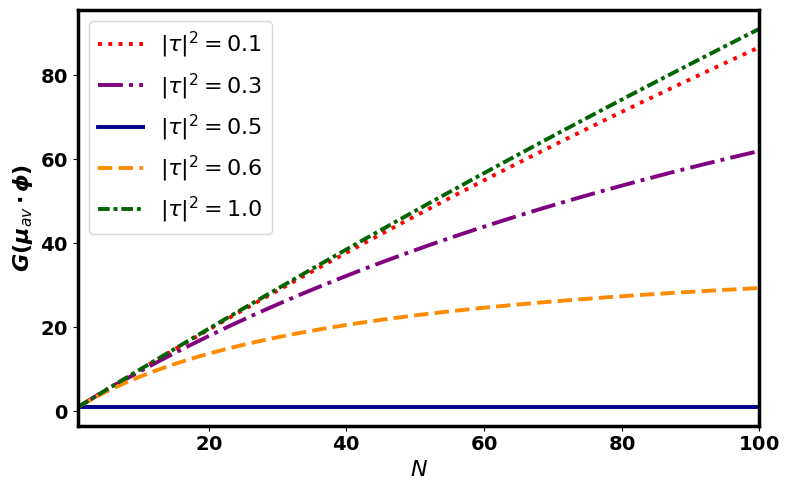}
        \label{b}
(b)
    \end{minipage}

     \caption{Sensitivity gain associated with the average phase estimation as a function of $N$, for $a=0.5$.  (a) Results for different values of $\overline{n}_T$ with $|\tau|^2 = 0.8$.  (b) Results for different values of $|\tau|^2$ with $\overline{n}_T = 100$.}  
     \label{gainav}     
\end{figure}
 In Fig.~\eqref{gainav}, we investigate the sensitivity gain of the entangled strategy over the separable strategy for average-phase estimation. Panel (a) shows the gain as a function of the number of interferometers $N$ for different values of the total average photon number $\overline{n}_T$. The gain increases with $N$ for all the considered values of $\overline{n}_T$. At low values of $\overline{n}_T$, the gain approaches a saturation regime as $N$
increases. In contrast, for larger values of $\overline{n}_T$, the gain exhibits an approximately linear scaling with $N$, particularly in the regime $\overline{n}_T \gg N$. This scaling demonstrates that increasing the available photon resources enhances the relative sensitivity
advantage of the entangled strategy.
Panel (b) shows the gain as a function of the transmission coefficient of the first beam splitter for different values of $N$. For the balanced configuration, the gain is equal to unity, indicating identical sensitivity for the entangled and separable strategies in average-phase estimation. As the beam-splitter configuration becomes increasingly unbalanced, the gain increases, indicating an enhancement of the sensitivity advantage provided by the entangled strategy. Furthermore,
for a fixed beam-splitter transmission, larger values of $N$ lead to a higher gain, showing that the relative advantage of the entangled strategy increases with the network size.
 
\subsection{Impact of photon losses}
\label{subsec:loss}
The analysis presented so far assumes an ideal lossless optical network. In realistic implementations, internal photon losses may affect the achievable phase sensitivity. Here, the losses in the two arms of each MZI are modeled as photon-loss channels, where the optical modes are coupled to independent environmental vacuum modes according to
\begin{equation}
\hat a'_k
\longrightarrow
\sqrt{\eta_2}\,\hat a'_k
+
\sqrt{1-\eta_2}\,\hat e_{a,k},
\qquad
\hat b'_k
\longrightarrow
\sqrt{\eta_1}\,\hat b'_k
+
\sqrt{1-\eta_1}\,\hat e_{b,k},
\label{loss_transformation}
\end{equation}
where $\hat e_{a,k}$ and $\hat e_{b,k}$ denote independent environmental vacuum modes. The parameters $0\leq\eta_1,\eta_2\leq1$ characterize the photon transmission through the $b'_k$ and $a'_k$ arms, respectively, and are assumed to be the same for all interferometers. The parameters $\eta_1$ and $\eta_2$ describe the internal losses and should not be confused with the beam-splitter coefficients $\tau$ and $r$. The lossless case is recovered for $\eta_1=\eta_2=1$.
For the mode-entangled strategy, the QCRB in the presence of internal
losses becomes
\begin{equation}
\Delta^2(\boldsymbol{\mu}\cdot\boldsymbol{\phi})_{Q}^{\mathrm{me}}
=
\frac{
\left(\sum_{k=1}^{N}|\mu_k|\right)^2
}{
\overline{n}_{\xi}
\left(\eta_1|\tau|^2+\eta_2|r|^2\right)
+
\dfrac{\overline{n}_{\xi}^{2}}{2a}
\left(\eta_1|\tau|^2-\eta_2|r|^2\right)^2
}.
\label{Qme_loss}
\end{equation}
For convenience, we define
\begin{equation}
\Lambda_{\eta}
=
\eta_1|\tau|^2-\eta_2|r|^2.
\label{loss_imbalance}
\end{equation}
In the high-photon-number regime, $\overline{n}_{\xi}\gg1$, and for $\Lambda_{\eta}\neq0$, the quadratic contribution dominates, and the QCRB becomes
\begin{equation}
\Delta^2(\boldsymbol{\mu}\cdot\boldsymbol{\phi})_{Q}^{\mathrm{me}}
\simeq
\frac{
2a\left(\sum_{k=1}^{N}|\mu_k|\right)^2
}{
\overline{n}_{\xi}^{2}\Lambda_{\eta}^{2}
}
\sim
\frac{1}{\overline{n}_{\xi}^{2}}.
\label{Qme_loss_asym}
\end{equation}
Hence, within the internal-loss model considered here, the Heisenberg-like scaling is preserved as long as $\Lambda_{\eta}\neq0$, although the attainable sensitivity is modified by the loss-dependent factor $\Lambda_{\eta}^{2}$. When $\Lambda_{\eta}=0$, i.e., $\eta_1|\tau|^2=\eta_2|r|^2$, the quadratic contribution vanishes and the scaling reduces to $1/\overline{n}_{\xi}$. In the lossless case, this condition corresponds to the balanced beam splitter, $|\tau|^2=|r|^2$, whereas unequal internal losses shift the cancellation point. The effect of internal losses on the relative advantage of the two strategies can be examined through the gain. As in the lossless case, for single-phase estimation the two strategies provide the same precision,
\begin{equation}
G^{(\mathrm{loss})}
(\boldsymbol{\mu}_{\mathrm{sng}}\cdot\boldsymbol{\phi})
=1.
\label{Gain_loss_single}
\end{equation}
For average-phase estimation, $\mu_k=1/N$, and under the same totalaverage-photon-number constraint, $\overline{n}_{\xi}=\overline{n}_{T} $, the gain becomes
\begin{equation}
G^{(\mathrm{loss})}
(\boldsymbol{\mu}_{\mathrm{av}}\cdot\boldsymbol{\phi})
=
\frac{
2a\left(\eta_1|\tau|^2+\eta_2|r|^2\right)
+
\overline{n}_{T}\Lambda_{\eta}^{2}
}{
2a\left(\eta_1|\tau|^2+\eta_2|r|^2\right)
+
\dfrac{\overline{n}_{T}}{N}\Lambda_{\eta}^{2}
}.
\label{Gain_loss_av}
\end{equation}
For $\overline{n}\gg1$ and $\Lambda_{\eta}\neq0$, the quadratic contributions dominate, yielding
\begin{equation}
\lim_{\overline{n}_{T}\rightarrow\infty}
G^{(\mathrm{loss})}
(\boldsymbol{\mu}_{\mathrm{av}}\cdot\boldsymbol{\phi})
=
N.
\label{Gain_loss_high}
\end{equation}
Thus, the asymptotic $N$-fold advantage of the mode-entangled strategy for average-phase estimation is retained under finite internal losses. At finite photon resources, however, increasing loss reduces the gain. For symmetric internal losses, $\eta_1=\eta_2=\eta$, the gain simplifies to
\begin{equation}
G^{(\mathrm{loss})}
(\boldsymbol{\mu}_{\mathrm{av}}\cdot\boldsymbol{\phi})
=
\frac{
2a+
\eta \overline{n}_{T}
\left(|\tau|^2-|r|^2\right)^2
}{
2a+
\dfrac{\eta\overline{n}_{T}}{N}
\left(|\tau|^2-|r|^2\right)^2
}.
\label{Gain_symmetric_loss}
\end{equation}
For a fixed finite photon resource, decreasing $\eta$, corresponding to increasing internal photon loss, reduces the relative contribution of the quadratic term and consequently decreases the gain. In the strong-loss limit, $\eta\rightarrow0$, the distinction between the two strategies is progressively suppressed and $G^{(\mathrm{loss})}\rightarrow1$. Conversely, for any fixed nonzero $\eta$ and away from the balanced beam-splitter condition, the high-photon-number limit still yields $G^{(\mathrm{loss})}\rightarrow N$. These two limits correspond to different physical regimes: increasing loss suppresses the mode-entangled advantage at fixed photon resources, whereas in the high-photon-number regime the surviving quadratic contribution can still dominate for any fixed nonzero value of $\eta$.

\section{Conclusion}

In this work, we have investigated a distributed quantum sensing protocol based on a configurable network of $N$ Mach--Zehnder interferometers driven by an SU(1,1) Perelomov coherent state. The estimation problem is formulated in terms of a general linear combination $\boldsymbol{\mu}\cdot\boldsymbol{\phi}$ of relative phase shifts, allowing both local and global sensing tasks to be addressed within the same multiparameter framework. Two sensing strategies have been analysed and compared under the same total average photon-number constraint: an entangled strategy, in which a single SU(1,1) Perelomov coherent state is distributed across the sensing modes through a passive quantum circuit, and a separable strategy, in which independent probe states are employed at the individual interferometers.

The analysis based on the QFIM and the QCRB highlights the combined role of the nonclassical input state and the quantum-circuit-based distribution of the sensing resources. The SU(1,1) Perelomov coherent state exhibits nonclassical photon-number fluctuations, including the quadratic contribution $\overline{n}_{\xi}^{\,2}/(2a)$, which provides an important resource for enhanced phase sensitivity. In the entangled strategy, the quantum circuit coherently distributes this resource across the sensor network and generates correlations between the sensing modes, which are reflected in the off-diagonal elements of the QFIM. Optimization of the quantum-circuit parameters further determines the photon allocation among the sensing modes according to the weighting
vector $\boldsymbol{\mu}$ defining the phase function to be estimated. Thus, the sensing performance results from the interplay between the
nonclassical properties of the probe state, the correlations established across the network, and the optimized distribution of the available
photon resources. Our analysis also reveals the important role of the transmission coefficient of the first beam splitter of each MZI. The balanced
configuration leads to shot-noise scaling, whereas optimization with respect to the first beam-splitter transmission favours highly unbalanced
configurations. Although conventional two-path interference is strongly suppressed in these limiting configurations, phase information remains
encoded at the quantum-state level. Owing to the quadratic photon-number fluctuations of the SU(1,1) Perelomov coherent state, the optimized QCRB
scales as $1/\overline{n}_{\xi}^{\,2}$ in the high-photon-number regime, corresponding to Heisenberg scaling. This result clarifies that the enhanced sensitivity predicted in the highly unbalanced regime originates from the quantum-statistical properties of the probe and its phase encoding rather than from conventional interference visibility. The comparison between the entangled and separable strategies further clarifies the conditions under which the quantum-circuit-based distribution provides a metrological advantage. For single-phase estimation, the gain reduces to unity, indicating identical sensitivity
for the two strategies. In contrast, the maximum gain is obtained for average-phase estimation, where the estimation task involves all sensing modes. This result shows that the advantage of the entangled strategy is particularly relevant for the estimation of global phase functions distributed across the sensor network. Moreover, the gain increases with the number of interferometers and the available photon resources and exhibits an approximately linear dependence on $N$ in the regime $\overline{n}_{T}\gg N$. The gain also increases as the first beam splitter becomes increasingly unbalanced. These results establish the parameter regimes in which the entangled strategy provides an advantage over its separable counterpart and demonstrate the relevance of the proposed architecture for distributed quantum sensing in large interferometer networks. The effect of internal photon losses has also been examined. Compared
with the lossless case, internal losses modify the attainable sensitivity and reduce the mode-entangled gain at finite photon resources. Nevertheless, within the loss model considered here, the mode-entangled strategy continues to exhibit the $1/\overline{n}_{\xi}^{2}$ asymptotic scaling for $\Lambda_{\eta}\neq0$, whereas $\Lambda_{\eta}=0$ leads to shot-noise scaling. For average-phase estimation, in the presence of finite internal losses, the mode-entangled gain still approaches $N$
in the high-photon-number regime for $\Lambda_{\eta}\neq0$. \\Several perspectives arise from the present work. Further developments include the implementation of explicit measurement and detection schemes and the assessment of the proposed protocol under realistic experimental conditions. The theoretical framework can also be generalized to broader input-state configurations, particularly by injecting multiple non-vacuum states into the quantum circuit instead of a single SU(1,1) Perelomov coherent state. In addition, the vacuum input port of each Mach--Zehnder interferometer can be replaced by a coherent or squeezed state, allowing the investigation of more general two-input interferometric configurations. These extensions provide relevant directions for broadening the scope of the proposed distributed quantum sensing framework.

\appendix
\section{Analytical Lower and Upper Bounds in a Separable Strategy}
 In this section, we give the demostration of the inequality and expressions~\eqref{mssng} and~\eqref{msav}.
The minimization of the Eq.~\eqref{QCRBms1} is achieved by applying the Cauchy-Schwarz inequality, given by,
 \begin{equation}
     \left( \sum_{k}^N \alpha_k \beta_k \right)^{2} \leq \left( \sum_{k}^N \alpha_k^{2} \right)\left( \sum_{k}^N \beta_k^{2} \right), 
    \quad \alpha_k,\; \beta_k \geq 0.
    \label{Csh1}
 \end{equation}
The equality  in \eqref{Csh1} holds if and only if
\begin{equation}
\alpha_k  = c \; \beta_k,
\label{cnd}
\end{equation} 
where \( c\) is a coefficient independent of the index \(k\). In our case the quantities \(\alpha_k\) and \(\beta_k\) are given
\begin{equation}
 \alpha_k = \frac{|\mu_k|}{\sqrt{\overline{n}_{\xi_{k}}+ \frac{\overline{n}_{\xi_{k}}^2}{2a} (\mathcal{\tau}^2-r^2)^2}}, \qquad
\beta_k = \sqrt{\overline{n}_{\xi_{k}}+ \frac{\overline{n}_{\xi_{k}}^2}{2a} (\mathcal{\tau}^2-r^2)^2},   
\end{equation}
which leads to
\begin{equation}
 \Delta^2(\mathbf{\mu} \cdot \boldsymbol{\phi})_{QCRB}^{ms} \geq \frac{ ( \sum_{k=1}^{N}\mu_{k})^{2}}{ \overline{n}_{\xi}+ \frac{ \sum_{k=1}^{N}\overline{n}_{\xi}^2}{2a} (\mathcal{\tau}^2-r^2)^2 },
 \label{QCRBms1}
\end{equation}
 We have 
\begin{equation}
\sum_{k=1}^{N} \overline{n}_{\xi_k}^{2} \; \leq \; \overline{n}_{T}^2 \; \leq \; N \sum_{k=1}^{N} \overline{n}_{\xi_k}^{2}, \qquad \overline{n}_{T} = \sum_{k=1}^{N} \overline{n}_{\xi_k}.
\label{enq}
\end{equation}
equality in \eqref{enq} achieved if and only if:
\begin{equation}
   \sum_{k=1}^{N} \overline{n}_{\xi_k}^{2} =
\begin{cases}
  \frac{\overline{n}_{T}^2}{N} & \text{if}\qquad \overline{n}_{\xi_k} =  \frac{\overline{n}_{T}}{N}  ,\\[2mm]
\overline{n}_{T}^2 & \text{if}\qquad \overline{n}_{\xi_l} = \overline{n}_{T}, \: \overline{n}_{\xi_k}= 0 \; \text{for} \; k\neq l.
\end{cases} 
\label{case}
\end{equation} 
 By substituting this optimal value $\overline{n}_{\xi_k} = \frac{\overline{n}_{T}}{N}$ into the saturation condition~\eqref{cnd}, we can show that the weighting coefficients must be identical:
\begin{equation}
\mu_k = \frac{1}{N}, \quad \text{for all}\quad k = 1, \ldots, N.
\end{equation}
This corresponds to the estimation of the average phase case discussed in the previous section.
Under these optimal conditions, we finally obtain the optimal Cramér-Rao bounds for the separable strategy:
\begin{equation}
\Delta^2(\boldsymbol{\mu_{av}}\!\cdot\!\boldsymbol{\phi})_{Q}^{ms}
=
\frac{1}
{\overline{n}_{T}
+
\frac{\overline{n}_{T}^{2}}{2Ma}(\mathcal{\tau}^2-r^2)^2}, 
\end{equation}
Similarly, for $\overline{n}_{\xi_l} = \overline{n}_{T}$, we can show,
\begin{equation}
|\mu_l| = \frac{1}{\sqrt{N}}, \quad |\mu_k|=0 \quad \text{for} \; k\neq l.
\end{equation}
corresponds to the estimation of a single phase discussed above. Then the optimal QCRB is given by 
\begin{equation}
\Delta^2(\boldsymbol{\mu_{sng}}\!\cdot\!\boldsymbol{\phi})_{Q}^{ms}
=
\frac{1}
{N(\overline{n}_{T}
+
\frac{\overline{n}_{T}^{2}}{2a}(\mathcal{\tau}^2-r^2)^2)},
\end{equation}
It is clear that $\Delta^2(\boldsymbol{\mu_{sng}}\!\cdot\!\boldsymbol{\phi})_{Q}^{ms} < \Delta^2(\boldsymbol{\mu_{av}}\!\cdot\!\boldsymbol{\phi})_{Q}^{ms}$. From equation \eqref{enq}, we deduce that, regardless of the linear combination of phases considered,  the strategy yields a value bounded by the two extreme configurations as the entangled strategy.
\begin{equation}
  \Delta^2(\boldsymbol{\mu_{sng}}\!\cdot\!\boldsymbol{\phi})_{Q}^{ms} \leq \Delta^2(\boldsymbol{\mu}\!\cdot\!\boldsymbol{\phi})_{Q}^{ms}\leq\Delta^2(\boldsymbol{\mu_{av}}\!\cdot\!\boldsymbol{\phi})_{Q}^{ms}.
\end{equation} 

\section*{Data availibility}
The datasets used and/or analysed during the current study available from the corresponding author on reasonable request.

\section*{Conflicts of Interest}
The authors declare no conflict of interest.

\end{document}